\documentclass[a4paper,oneside,12pt]{article}%
\usepackage[a4paper]{geometry}
\usepackage{graphicx}
\usepackage{amsmath}
\usepackage{amsfonts}
\usepackage{amssymb}%
\providecommand{\U}[1]{\protect\rule{.1in}{.1in}}

\begin{document}

\title{Can an \textquotedblleft impulse response\textquotedblright\ really be defined
for a photoreceiver?}
\author{F. Javier Fraile-Pelaez
\and {\small Dept. de Teor\'{\i}a de la Se\~{n}al y Comunicaciones, Universidad de
Vigo,}
\and {\small ETS Ingenieros de Telecomunicaci\'{o}n, Campus Universitario E-36310
Vigo (Spain).}}
\date{}
\maketitle

\begin{abstract}
In this paper we examine the validity of the concept of impulse response
employed to characterize the time response and the signal-to-noise ratio of
p-i-n and similar photodetecting devices. We analyze critically the way in
which the formalism of analog linear systems has been extrapolated, by
employing results from macroscopic electromagnetic theory such as the
Shockley--Ramo theorem or any equivalent approach, to the extreme case of a
single-photon detection. We argue that the concept of \textquotedblleft
response to an optical impulse\textquotedblright\ is ill-defined in the
customary terms it is envisioned in the literature, this is, as an output
current pulse having a certain predictable, calculated temporal shape, in
response to the detection of an optical \textquotedblleft Dirac
delta\textquotedblright\ impulse, conceived in turn as the absorption of a
single photon.

\end{abstract}

\section{Introduction}

It is well known that the ultimate sensitivity of any photodetector is
determined by the quantum noise of the radiation itself. Specifically for a
photodiode receiver, this means that the noise present at the output current
of the diode should be an exact reproduction\ of the intrinsic noise of the
impinging radiation, with no any other excess noise contributions. Obviously,
the current-voltage amplification at the electronic stage of the receiver
should also be noiseless, consistent with the \textquotedblleft ideal
receiver\textquotedblright\ assumption. In other words, once all additional
(electronic) noise sources have been removed, the process of noiseless
photodetection amounts to photon-counting through ideally equivalent temporal electron-counting.

On the other hand, the functional modelling of a photoreceiver systematically
makes use of an essential concept taken from linear systems theory: the
\textquotedblleft impulse response of the receiver,\textquotedblright%
\ \cite{senhal} which is required for the analysis of both signal and noise
performance of any linear, invariant system. For an analog system, the impulse
response is defined as the output time signal when the input is an
instantaneous impulse of unit area, i.e. a Dirac delta, $\delta(t),$ which
contains all frequencies homogeneously, from $0$ to $\infty$. Such an impulse
is unrealizable (and surely unphysical), but its mathematical usefulness makes
it convenient to assume its existence, at least in the approximate form of a
physical impulse having a duration much shorter than any characteristic time
of the system. Thus, in the case of an electrical circuit, one can think of a
delta-like impulse of voltage, for example. In the case of incoherent optical
reception, the input \textquotedblleft signal\textquotedblright\ is the
time-varying optical power $P(t),$ so the input impulse is to be described
mathematically as $P(t)=\delta(t).$

It should be kept in mind that all signals are inherently analog in this
formalism. Actually, to a great extent, the Dirac delta works as an \emph{ad
hoc} artifact intended to allow hypothetical point-like objects (masses,
charges...) to \textquotedblleft live\textquotedblright\ in continuous spatial
or temporal domains, which would otherwise be unconceivable; if space-time is
thought continuous, at least differential intervals are needed to contain a
non-null amount of any magnitude, since a discrete \emph{point} is, in
mathematical terms, a zero measure set, thus meaningless. Only if one accepts
that a space or time point can accommodate a \textquotedblleft Dirac
delta\textquotedblright\ (of charge, say), can the problem be skipped.

The above considerations lead us to the following point. Consider the optical
signal to be a narrow-band modulated optical flux $\bar{Q}(t)=P(t)/(h\nu)$
(photons/s), where the overbar denotes statistical averaging and $\nu$ is the
central optical frequency. This corresponds well to the archetypical case of a
laser (or even LED) beam modulated in intensity by a low frequency (baseband,
RF, microwave) signal varying like $P(t).$ Contrary to what is frequently
implied in the literature, the \textquotedblleft unit\textquotedblright%
\ impulse at the input of the detector is \emph{not} \textquotedblleft one
photon\textquotedblright---in spite of the cardinal number. This confusion,
detected in many textbook presentations, arises surely from the fact that the
electromagnetic field, roughly speaking, happens to be quantized \emph{in
amplitude}, whereas the Dirac delta formalism was never intended to deal with
\textquotedblleft quantized analog\textquotedblright\ signals---incidentally,
a concept which does not exist in linear systems theory.

As far as the signal part of the signal and noise calculations is concerned,
the problem can be surmounted easily for two related reasons. First, the unit
amplitude of the Dirac delta is purely conventional and without consequences
in a linear system; obviously, if the impulse $A\delta(t)$ is employed at the
system input, the system response to the unit impulse will merely be the
actual output divided by $A.$ In other words, it is the temporal condensation
that matters, not the amplitude. Second, in view of the previous
consideration, any sufficiently short optical pulse, yet simultaneously
intense enough to clear up any concern on signal level quantization, will be a
perfectly valid approximation to the unit input impulse.

Things change when the focus is put on the noise. Particularly in the case of
the photonic signal noise, the inherent amplitude quantization cannot be
disguised anymore and the solution described above is unfeasible. One thus has
to confront a frequently overlooked issue which threatens the typical
automatic extension of the linear system formalism to handle noise of quantum
origin. In Sections 2 and 3 we review, very briefly, the standard theory of
the signal noise as routinely applied to the linear system model of
photoreceiver. The problems carried out by the accepted formalism are
discussed in Section 4. Section 5 contains the conclusions.

\section{Optical shot noise in the photoreceiver model}

Quantum noise is almost synonymous of shot noise as far as a photoreceiver is
concerned. Only two or three simple statistical concepts are needed to
describe the photodetection process in the simple fashion in which it is
usually modelled, and a straightforward correspondence can be apparently
established between the mathematical route and the physical route. Thus,
assuming a coherent light source, the random arrival times of the photons are
governed by a Poisson distribution characterized by its average $\bar{N}$,
related in turn to the average rate of the photon flux through $\bar{N}%
=\bar{Q}T$, with $T$ the \textquotedblleft photocounting\textquotedblright%
\ period. One could anticipate at this point that $T$ will be roughly equal to
the inverse of the bandwidth, which is basically true.

\begin{figure}
[h]
\begin{center}
\includegraphics[
height=1.6779in,
width=2.0855in
]%
{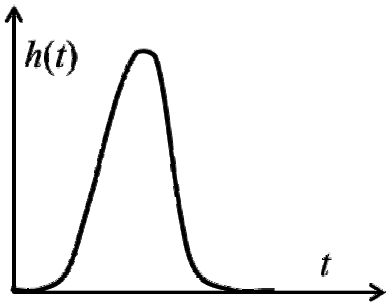}%
\caption[When applied to a photodetector, the formalism of linear systems
seeks to calculate the \textquotedblleft impulse response\textquotedblright%
\ as the photocurrent pulse at the output of the device which corresponds to
the detection of just one photon. In absence of internal gain, such elementary
current pulse is predicted to have the form $i(t)=qh(t),$ with $q$ the
electronic charge and $h(t)$ the pulse shape, satisfying $\int_{-\infty
}^{\infty}h(t)dt=1.$]{When applied to a photodetector, the formalism of linear
systems seeks to calculate the \textquotedblleft impulse
response\textquotedblright\ as the photocurrent pulse at the output of the
device which corresponds to the detection of just one photon. In absence of
internal gain, such elementary current pulse is predicted to have the form
$i(t)=qh(t),$ with $q$ the electronic charge and $h(t)$ the pulse shape,
satisfying $\int_{-\infty}^{\infty}h(t)dt=1.$}%
\label{fig1}%
\end{center}
\end{figure}

Next, as an optical-electrical transducer, the photodetector transmutes the
photon absorptions into charge carriers---the mathematical consequence being a
mere multiplication of the actual instantaneous photon flux, $Q(t)=%
{\textstyle\sum_{k}}
\delta(t-t_{k}),$ by the electron charge $q$ to arrive at the same delta train
function, but this time as an electrical current rather than a photon flux:
$i_{\delta}(t)=q%
{\textstyle\sum_{k}}
\delta(t-t_{k}).$ Certainly, this impossible current is only a conceptual
intermediate step toward the \textquotedblleft real\textquotedblright%
\ current, which in general is described by the expression%

\begin{equation}
i(t)=q%
{\textstyle\sum_{k}}
M_{k}h_{k}(t-t_{k}), \label{1}%
\end{equation}
where $h_{k}(t)$ is the shape of the \emph{current pulse} generated across the
terminals of the photodiode by the $k$-th absorbed photon. The prefactor
$M_{k}$ accounts for the possibility of the detector being an avalanche
photodiode (APD) with average gain $\bar{M},$ while the subindex $k$ of
$h_{k}$ reflects the fact that, even in a p-i-n photodiode, the shape of
current pulse will vary depending on the specific location within the
photodiode where the photon has been absorbed \cite{Fraile}. Expression
(\ref{1}) is most often oversimplified by ignoring the random character of
$h_{k}$ and writing a fixed $h(t)$, sketched in Fig. \ref{fig1}, which is then
identified with the \textquotedblleft impulse response\textquotedblright\ of
the linear system, its Fourier transform $H(\omega)$ being the photodetector
transfer function.%

Considering the specificities of an APD is unnecessary for the purpose of the
present discussion, so we will take $M_{k}=1$ and focus on a p-i-n photodiode.
The shape of $h_{k}(t)$ is determined by the geometry and structure of the
diode, mainly the width of the intrinsic layer. Fig. 1 sketches the form of
the current pulse, which---always within the frame of the described
approach---arises from the transit of \emph{one} electron-hole pair
photogenerated (typically) somewhere in the space charge region, toward the
positive and negative, respectively, electrodes of the structure. These
transit times determine the ultimate bandwidth of the photodetector.

\section{Impulse response and sub-electron charge}

The area of any elementary current pulse as described above is given by%

\begin{equation}
\int_{-\infty}^{\infty}i(t)dt=q\int_{-\infty}^{\infty}h_{k}(t)dt=q, \label{2}%
\end{equation}
which manifests the transfer of one electron charge during the duration of the
pulse, or, expressed more accurately, the passage of a total charge $q$ across
an imaginary plane located at any point along the electrical circuit. Thus, at
the end of the \textquotedblleft flight time\textquotedblright\ of the
electron and the hole, assuming they do not recombine before being collected
at the electrodes, one can safely say that a total charge of one electron has
moved, as a conduction current, along the whole circuit. However, expression
(\ref{1}) has a very discomforting feature. If $h_{k}(t)$ is truly a current
shape and the actual pulse duration spreads, say, from $t=0$ to $t=T_{p},$ one
should be able to observe a \emph{fractional} charge $q_{f}$ given by%

\begin{equation}
q_{f}=q\int_{t_{1}}^{t_{2}}h_{k}(t)dt \label{3}%
\end{equation}
during any finite interval $[t_{1},t_{2}],$ with $0\leq t_{1}<t_{2}\leq
T_{p}.$ However striking this consequence of the formalism may look, seemingly
it has never deserved a remark in any textbook or article, passing completely
unnoticed in the literature to the author's knowledge.

It is necessary to recall the origin of the theory leading to this somewhat
stunning result (\ref{3}). Essentially, this is the Ramo or Shockley--Ramo
theorem (SRT) \cite{S}, \cite{R}, first applied to determine the shape of the
anode current of a vacuum tube by computing the charge electrostatically
induced on the plate during the \textquotedblleft flight\textquotedblright\ of
the electronic space charge across the inter-electrode space. The SRT has been
used intensively and extended to deal with other scenarios, including solid
state devices (see, for example, \cite{vali}, \cite{ext}).

Ramo's original proof basically appeals to the energy balance, provides a
relatively simple result which facilitates otherwise more cumbersome
calculations. However, for the purpose of the present discussion, we will use
an argument based directly on the Maxwell equations since, with the toy model
to be used here---also frequently employed in the literature---, both
approaches have about the same simplicity while the latter provides some more
physical insight.

Figure 2 illustrates, with a very simple scheme, how the computation of the
\textquotedblleft impulse current\textquotedblright\ is almost universally
made. To focus on the essential concepts, we will consider a typical
one-dimensional, homogeneous structure of dielectric constant $\varepsilon$
(it could equally be vacuum) bounded by two conducting planes. This could
represent, for example, the intrinsic zone of a p-i-n photodiode sandwiched
between p$^{+}$ and n$^{-}$ zones. As assumed in the ideal linear model, the
voltage $V$ across the dielectric remains constant, regardless of the
photogenerated space charge and current. The \textquotedblleft unit
impulse\textquotedblright\ will then be materialized by the instantaneous
absorption of one photon anywhere between the electrodes, say at $x=x_{0},$
and the corresponding generation of a single charge or an electron-hole pair
at that point. Assume, for the sake of maximum simplicity, that just one
electron is photogenerated. This \textquotedblleft
one-dimensional\textquotedblright\ electron will be described as a discrete
charge sheet on the plane located at $x_{0},$ with a surface density charge
$\sigma_{e}$ such that $\sigma_{e}A=-q,$ with $-q$ the electronic charge and
$A$ the transversal area considered. It should be noted, at this point, that
we do accept the one-dimensional modelization of the electronic charge, merely
for obvious reasons of mathematical convenience, and this has nothing to do
with the conceptual difficulties that are the object of this article.%

\begin{figure}
[bh]
\begin{center}
\includegraphics[
height=2.4085in,
width=3.105in
]%
{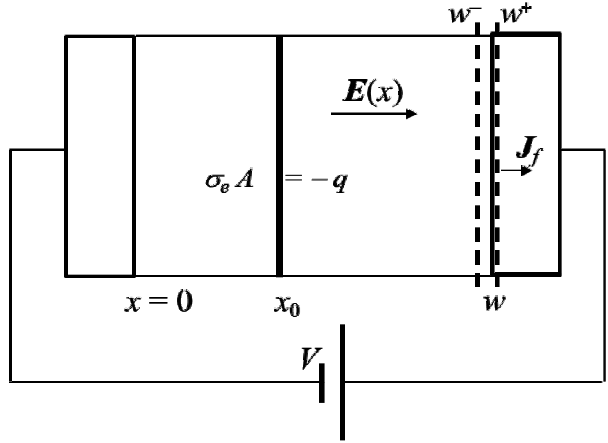}%
\caption{One-dimensional model illustrating the way in which the photocurrent
pulse corresponding to the detection of one photon is obtained.}%
\end{center}
\end{figure}

The electron generated at $x_{0}$ will then move toward the positive electrode
at $z=d$ under the influence of the bias field created by $V_{b}.$ As is
customarily made, we keep it simple and assume that it moves at a constant
saturation velocity $v_{s}.$ The argument now is that, during the whole flight
time of the electron, the time-dependent electric field it generates will
electrostatically induce a continuously-varying current flowing through the
electrodes at $x=0$ and $x=w,$ thus resulting in the circulation of a short
current pulse along the circuit. This should be the \textquotedblleft impulse
response\textquotedblright\ of the photodetector.

To calculate the shape of the aforementioned current pulse, we make use of the
law of the conservation of charge, which follows readily from the Maxwell
equations and reads, for the \textit{free} charge and current density,
$\nabla\cdot\boldsymbol{J}_{f}=-\partial\rho_{f}/\partial t.$ Considering a
certain volume $V$ and applying the Gauss theorem, we obtain the integral
relation $%
{\textstyle\oint}
{}_{S}\boldsymbol{J}_{f}\cdot d\boldsymbol{S}=-\partial Q_{f}/\partial t,$
with $Q_{f}=\int_{V}\rho_{f}dV,$ the total free charge enclosed. We choose to
use a rectangular Gaussian box limited by the planes $x=w^{-}$ (very close to
the right-hand plate surface) and $x=w^{+}$ (just inside the right-hand
plate). The electric field vector is defined as $\boldsymbol{E}%
(x)=E(x)\,\boldsymbol{\hat{x}}.$ In the surface integral, $d\boldsymbol{S}%
=\pm\boldsymbol{\hat{x}}dS$ on the right/left plane and $-\boldsymbol{\hat{x}%
}dS$ on the left plane, while $\boldsymbol{J}_{f}=J_{f}\,\boldsymbol{\hat{x}%
}.$Thus, $%
{\textstyle\oint}
{}_{S}\boldsymbol{J}_{f}\cdot d\boldsymbol{S}=-J_{f}(w^{-},t)A+J_{f}%
(w^{+},t)A.$ There is no free current at $w^{-},$ so $J(w^{-},t)=0.$ The total
free charge inside the volume is given by $Q_{f}=\sigma_{w}A,$ with
$\sigma_{w}$ the surface charge density at the right-hand plate, which can be
related to the normal field at the conductor surface by the equation
$\boldsymbol{E}(w,t)=-\boldsymbol{\hat{x}}\sigma_{w}(t)/\varepsilon$ in the
quasi-static approximation, so one finally obtains $J_{f}(w^{+},t)=\varepsilon
\partial E(w,t)/\partial t.$ The field $E(w,t),$ determined by the moving
electron sheet charge $\sigma_{e}$, can be computed using the Gauss law; it is
easy, in this elementary case, to arrive at the result%

\begin{equation}
E(w,t)=-\frac{V}{w}+\frac{x_{0}(t)}{w}\frac{\sigma_{e}}{\varepsilon}.
\end{equation}
Noting that $dx_{0}(t)/dt=v_{s},$ the result $J_{f}(w^{+},t)=\sigma_{e}%
v_{s}/w$ follows. The total plate current is given by%

\begin{equation}
I(t)=\sigma_{e}\frac{v_{s}}{w}A=-q\frac{v_{s}}{w}\text{\qquad during
}0<t<(w-x_{0})/v_{s}%
\end{equation}
(assuming the photocarrier is generated at $t=0$ and instantaneously
accelerated to $v_{s}$). The total charge crossing the $x=w$ plane from left
to right during the time interval $[0,(w-x_{0})/v_{s}]$ will be%

\begin{equation}
q_{T}=\int_{0}^{(w-x_{0})/v_{s}}I(t)dt=-\frac{q}{w}\left(  w-x_{0}\right)  ,
\end{equation}
which is smaller than the charge of single electron. Actually, it is the
photogenerated hole we have disregarded which provides, with a similarly
shaped current pulse, the remaining charge, so that the total value $-q$ is obtained.

To summarize, we see that the formalism predicts a current pulse of
rectangular shape (assuming equal electron and hole velocities) containing a
total charge equal to the electron charge. Further refinements in the model,
such as field-dependent carrier drift velocities and others, would lead to
more or less complicated calculations and different resulting pulse shapes,
but the issue of the sub-electron charge persists as all models are developed
along the same key conceptual lines.

\section{Discussion}

As it is well known, when Maxwell's classical equations are applied to
material media, it is with the understanding that some process of macroscopic
averaging is necessarily carried out (a comprehensive study can be found in
\cite{robin}, for example). Of course, in a microscopic view, one can consider
individual charges formally described by a volume charge density of the type
$\rho(\boldsymbol{r})=\sum_{j}q_{j}\delta\lbrack\boldsymbol{r}-\boldsymbol{r}%
_{j}(t)],$ and still employ the corresponding spatially-continuous
electromagnetic fields consistently calculated. However, one cannot reasonably
expect that both approaches can be freely mixed in the same equations. When
the electrostatic normal field on a conducting surface is written as
$\boldsymbol{E}(\boldsymbol{r})=\boldsymbol{\hat{n}\,}\sigma(\boldsymbol{r}%
)/\varepsilon,$ it has been tacitly accepted that $\sigma$ is a mathematically
continuous function of $\boldsymbol{r},$ which can vary over $\boldsymbol{r}$
as smoothly as $\boldsymbol{E}$ (which \emph{is} a true continuous magnitude)
demands. Physically, this is obviously never the case, so an implicit
approximation is always under the rug;\ in this example, the approximation is
that---contrary to the single, discrete photogenerated carrier---a very thin
layer of atoms or molecules right below the surface contains so many free
electrons, that it can be macroscopically treated as a continuous surface
charge $\sigma.$ Of course, this view breaks down if one looks at the surface
too closely, but this limitation simply sets a scale limit beyond which it is
recognized that the details (usually not needed) will be missed.

The problem in our case is of a different nature. The continuous variation of
the field amplitude at $x=w,$ determined by the instantaneous position of the
photoelectron position during its \textquotedblleft flight\textquotedblright,
demands a continuously-varying surface charge density at the right-hand plate,
which, according to the charge conservation law, should mandatorily result in
a continuously-varying current density coming out of the $x=w$ plane. But of
course, all the steps in such a derivation take for granted that the
magnitudes involved are continuous or can be thought of as continuous with
sufficient accuracy. As remarked in the previous paragraph, this premise is
justified when \emph{many} microscopic entities can be averaged out into a
\textquotedblleft continuous\textquotedblright\ fluid, but the averaging
process becomes senseless when the set of (indivisible) microscopic entities
to be averaged happens to contain just one. In a way, proceeding in this
manner amounts to using a sort of a circular argument.

On the other hand, it is undeniable that the displacement of the single
photoelectron has to affect in some way the charges in the conducting left and
right plates. What is by no means obvious, is the automatic, seemingly
thoughtless assumption that the effect should be formalized through the same
standard approach which is used macroscopically. In contrast, if rather than
one single photogenerated electron located at point $\boldsymbol{r},$ there
were a small charge contribution $\Delta Q(\boldsymbol{r},t)=\rho
(\boldsymbol{r},t)\Delta V$ within a \textquotedblleft
differential\textquotedblright\ volume $\Delta V$ around $\boldsymbol{r}$
(which would be actually comprised of, say, millions of electrons), the
formalism could then be applied straightforwardly, since a temporal fraction
(say, $\alpha<1)$ of the output current pulse would still contain
$\alpha\Delta Q(\boldsymbol{r},t)/q$ electrons, hopefully a number large
enough to clear off any possible concern on charge discreteness.

Surprising as it may seem, the type of concerns expressed here never seem to
have drawn any interest in the specialized literature through the years. It is
futile to try to cite examples here, whether a few or many, since \emph{all}
known references could certainly be listed. With the sole purpose of not
leaving this point without any bibliographic support, we will mention, for
example, \cite{pho}--\cite{bowers} (references chosen almost at random with no
intention at all to single them out as being precedential or more
\textquotedblleft original\textquotedblright\ than others; the idea of the
\textquotedblleft single-photon\textquotedblright%
\ current-impulse---reluctantly assumed even in \cite{Fraile}---has propagated
through the photonics literature as a routine, to the point that it is
impossible to trace back its diffusion). Significantly, the first mentions to
a similar problem have only appeared in the relatively new field of mesoscopic
devices, as we exemplify next.

A single-electron transistor (SET) is a device where the so-called
\emph{Coulomb blockade} takes place \cite{hanson}. This process involves the
tunneling of individual electrons across a thin insulating barrier between two
conducting electrodes. Among other conditions, the theoretical model of the
Coulomb blockade requires a \emph{continuous} charge transfer from an external
source to the electrode. In this case, the conceptual problem posed by the
necessity of considering a continuous charge has not passed unnoticed, and
consensus seems to have been reached in recognizing that it is a continuous
\emph{spatial} displacement of the electronic charges around the atomic nuclei
of the metal, during the intervals between tunneling events, that may provide
the necessary \textquotedblleft continuous charge\textquotedblright. We
reproduce next a few, more or less similar, typical statements which can found
in the literature in regards with this issue. It will not go unnoticed, in any
case, that they all tend to be rather qualitative.

In \cite{tesis}, it is remarked that \emph{\textquotedblleft(...) the current
is determined by the current transferred through the conductor. Surprisingly
this transferred charge can have practically any value, in particular, a
fraction of the charge of a single electron. Hence, it is not quantized. This,
at first glance counterintuitive fact, is a consequence of the displacement of
the electron cloud against the lattice of atoms. This shift can be changed
continuously and thus the transferred charge is a continuous
quantity.\textquotedblright\ }Under the section title \textquotedblleft
Continuous charge transfer,\textquotedblright\ the following statement can be
found in \cite{otro}: \emph{\textquotedblleft(...) }$q$\emph{\ is not
necessarily a charge transferred through some imaginary cross-section of the
current leads (...) }$q$\emph{\ is rather }defined \emph{by the equation
}$U=q^{2}/2C$\emph{\ for the electrostatic energy of the capacitor, so that
}$q$\emph{\ is the }net \emph{surface charge of its
electrodes.\textquotedblright\ }Or, in \cite{har}, \emph{\textquotedblleft in
the macroscopic metallic leads ending at the barrier the electrons are in
extended states, i.e. they can move freely. Consequently the accumulated
charges }$+Q$\emph{\ and }$-Q$\emph{\ effectively result from a shift in the
average positions of the electrons on the two sides of the
barrier.\textquotedblright} As a final example, we quote the claim in
\cite{urb} that \emph{\textquotedblleft charge flow in a metal or a
semiconductor is a continuous process because conduction electrons are not
localized at specific positions. They form a quantum fluid which can be
shifted by an arbitrary small amount.\textquotedblright} Even if the
connection between the non-localizability of the electrons (of statistical
nature) and the electron flow being a \textquotedblleft continuous
process\textquotedblright\ appears somewhat obscure, the authors seem in any
case to appeal again to the continuous spatial displacement of the charges to
justify the formalism.

\section{Conclusions}

In view of all the previous considerations, we must finally decide whether it
is reasonable or not to expect that, upon absorption of just \emph{one} photon
at a specific point, a p-i-n photodetector (or any similar device) will really
be able to provide a photocurrent pulse having a continuous, \emph{repeatable}
shape $h(t),$ which can be calculated \emph{using the formalism} summarized
above. It is important to make precise what is meant by this: one should be
able to observe perfectly, after suitable amplification, this current pulse
shape on the screen of an oscilloscope, say. [Realistically, some additional
noise is to be expected, due to electronic components or multiplicative noise,
as in a APD photodetector or a photomultiplier tube, but this should certainly
not affect the alleged tangibility of a---perhaps small albeit
macroscopic---amplified output current instantaneously following the
continuous functional form $h(t)$]. Only the occurrence of this, and not any
other situation whatsoever, would fulfill the precise definition of
\textquotedblleft impulse response\textquotedblright\ of the photoreceiver as
a linear system, thus justifying the standard formalism under discussion.

It is interesting to note that, in all the descriptive accounts of the
photodetection process in a p-i-n photodiode, the argument provided to
calculate the quantum efficiency $\eta$ \emph{excludes} the photogenerated
carriers that recombine before reaching the corresponding electrode from the
current contribution. Indeed, the same reasoning is applied to compute the
efficiency of solar cells. There obviously underlies the idea than only the
discrete electrons or holes effectively \emph{collected} at the electrodes
will contribute to the output photocurrent. Actually if, as discussed with
regards to the Coulomb blockade in the previous section, the problematic
\textquotedblleft sub-electron\textquotedblright\ charge is really attributed
to small spatial displacements of the free electrons around the atomic nuclei
in the conductors, when the polarizing electric field causing this
displacement ceases to exist, i.e. when a traveling photoelectron, for
example, suddenly disappears by recombination before it reaches the positive
contact, the corresponding electronic charges should simply shift back to
their original positions, with no final consequences for the circuit current.
Obviously, one-photon detection has existed for a long time, but a survey of
the technical literature will show that virtually any approach to the subject
ends up considering basically a process of photocounting, thus
electron-counting; to the author's knowledge, no true experimental work
clearly undertaking the verification of the elusive $h(t)$ can be found in the literature.

To summarize, within the linear system description of a photoreceiver, the
impulse response of the system can be taken as the output voltage/current of
the receiver (which will follow, properly amplified, the temporal shape of the
device photocurrent) when a very short, (\textquotedblleft
delta\textquotedblright) impulse of photon flux is applied, with the proviso
that a sufficiently high number of photons \textquotedblleft
instantaneously\textquotedblright\ fill up the absorbing volume of the device.
Enough electrons/holes will then be created and drifted toward the terminal
electrodes forming a quasi-continuous current to which the familiar,
macroscopic model can be applied safely. This formalism will be valid for the
calculation of both the signal and the signal-to-noise ratio with
\emph{additive} noise (i.e., electronic noise: noise in the amplifier
circuitry, etc.) Indeed, it will also be valid for the quantum signal noise
itself, as long as the signal level is not so low that the corpuscular
character of the moving photocarriers becomes relevant, in the sense discussed
here. For the latter case, the mechanical extrapolation of the SRT or any
similar formalism to deal with the problem does not seem to be a very careful
decision, to say the least; we could express it stating that the equations
\textquotedblleft have been pushed too far\textquotedblright\ and nobody seems
to care... The idea that (even if the single incoming photon were to always be
absorbed at the same location) a true, deterministic analog linear-system
\textquotedblleft impulse response\textquotedblright\ can be conceived for
such an extreme situation, appears unrealistic, lacks any convincing
theoretical or experimental support, and should therefore be abandoned.

\end{document}